\documentclass[journal]{IEEEtran}
\usepackage{cite}
\usepackage{svg}
\ifCLASSINFOpdf
\else
\fi
\usepackage{amsmath}
\usepackage{algorithmic}
\usepackage{array}
  \usepackage{subfigure}
\usepackage{url}
\usepackage{pdfpages}
\usepackage{tikz}
\usetikzlibrary{arrows,decorations,backgrounds,shadows,plotmarks, positioning, calc,shapes, patterns,chains,intersections,external}
\usepackage[american,cuteinductors,smartlabels,EFvoltages]{circuitikz}

 \usepackage{xcolor}
	\definecolor{UCLgreenD}{rgb}{0.333,0.314,0.145}    % Dark 
	\definecolor{UCLgreenM}{rgb}{0.561,0.600,0.243}    % Medium
	\definecolor{UCLgreenB}{rgb}{0.710,0.741,0.000}    % Bright
	\definecolor{UCLgreenL}{rgb}{0.733,0.773,0.573}    % Light
	\definecolor{UCLredD}{rgb}{0.396,0.114,0.196}    % Dark 
	\definecolor{UCLredM}{rgb}{0.576,0.153,0.173}    % Medium
	\definecolor{UCLredB}{rgb}{0.835,0.000,0.196}    % Bright
	\definecolor{UCLredL}{rgb}{0.878,0.235,0.192}    % Light
	\definecolor{UCLpurpleD}{rgb}{0.294,0.220,0.298}       % Dark 
	\definecolor{UCLpurpleM}{rgb}{0.314,0.027,0.471}       % Medium
	\definecolor{UCLpurpleB}{rgb}{0.675,0.078,0.353}       % Bright (Officially: Bright Pink)
	\definecolor{UCLpurpleL}{rgb}{0.7764,0.690,0.737}      % Light
	\definecolor{UCLblueD}{RGB}{0,61,76}        % Dark 
	\definecolor{UCLblueM}{RGB}{0,40,85}        % Medium
	\definecolor{UCLblueB}{RGB}{0,151,169}      % Bright
	\definecolor{UCLblueL}{RGB}{141,185,202}    % Light
	\definecolor{UCLblueC}{RGB}{164,219,232}    % Celeste
	\definecolor{UCLblueIOE}{RGB}{50,85,164}    % Institute of Education
	\definecolor{UCLyellow}{RGB}{246,190,0}     
	\definecolor{UCLorange}{RGB}{234,118,0}   
	\definecolor{UCLgrey}{RGB}{140,130,121}   
	\definecolor{UCLbrown}{RGB}{78,54,41}  

\begin{document}
\title{A DC-Autotransformer based Multilevel Inverter for Automotive Applications}
\author{Ferdinand~Grimm,~John~Wood~
        and~Mehdi~Baghdadi%
%\author{Michael~Shell,~\IEEEmembership{Member,~IEEE,}
%        John~Doe,~\IEEEmembership{Fellow,~OSA,}
%        and~Jane~Doe,~\IEEEmembership{Life~Fellow,~IEEE}% <-this % stops a space
\thanks{F. Grimm and M. Baghdadi are with the Department
of Mechanical Engineering, University College London, London,
UK.}% <-this % stops a space
\thanks{J. Wood is with Silicon Contact LTD.}}% <-this % stops a space
\maketitle

\begin{abstract}
This paper proposes a novel multilevel inverter for automotive applications.
The topology consists of a modular DC-DC converter and a tap selector,
where the DC-DC converter provides several DC-output levels and the tap selector produces an AC signal by choosing different DC-output signals from the DC-DC converter. 
To produce the DC-levels, the DC-DC converter consists of a modular structure where the modules are connected in series.
The novelty is that the modules are connected both, magnetically in the AC-domain and electrically in the DC-domain. 
Due to the usage of low power switches in the modules, the proposed structure provides high efficiency.
Furthermore, the DC-DC converter is capable of self-balancing its modules and thus does not require large capacitors which yields a high power density. 
A prototype of the proposed converter is built and simulation, as well as experimental results, are used to verify the findings.
\end{abstract}

\begin{IEEEkeywords}
Multi-active bridge, autotransformer, multilevel converter, automotive inverter.
\end{IEEEkeywords}

% \fi
\IEEEpeerreviewmaketitle

\section{Introduction}
%%%%%%%%%%%%%%%%%%%%%%%%%%%%%%%%%%%%%%%% Introduction on DC/AC converters
%%% 1 MMC general
\IEEEPARstart{T}{he} modular multilevel converter (MMC) is a promising topology due to its low losses and high output signal resolution \cite{MMC1,MultiOverview6,MMC12,MMC13}.
Being comprised of many cells, it allows the usage of lower voltage ratings compared to conventional H-bridges \cite{MMC12}. 
To keep those voltages balanced, however large capacitors \cite{MultiOverview6} as well as complicated control \cite{MMC13} is required to allow the MMC to stay fully operational.
\newline
% 2. Dual active bridge
Most commonly, the dual active bridge \cite{DAB1,DAB2,DAB3} topology is used for the cells of the MMC.
Since its introduction in \cite{DAB1}, the dual active bridge has become a popular converter in automotive applications and its power electronic properties have been thoroughly studied in \cite{DAB2,DAB3}.
\newline
% 3. Dual active bridge
One of the distinctive features of the dual active bridge converter is that it provides two terminals.
The multi-active bridge is an extension of the dual active bridge that connects several active bridges through a multi-winding transformer
\cite{DAB_MultiAB,DAB_MultiDAB2,DAB_MultiDAB3,DAB_MultiDABAPP}.
This concept has been proposed in \cite{DAB_MultiAB}.
In this converter, each winding is connected to an active bridge and the energy is transferred magnetically over the transformer
Applications of the multi-active bridge converters can be found in the fields of grid connections \cite{DAB_MultiDABAPP} and electric aircraft \cite{DAB_MultiDAB2}.
\newline
%% 3 DC-autotransformer and applications
While the connection between submodules for conventional MMCs is entirely electric, the multi-active bridge topology allows implementing a connection that has both, electric and magnetic components
\cite{GBC1,GBC2,GBC3,GBC4}.
An interesting converter topology has been proposed at the Google  little box challenge and was published in \cite{GBC1}. 
Contrary to the previous topologies where several dual-active bridges were connected, this topology has all cells connected through magnetic coupling.
The outputs of the multi-active bridge are connected to a tap selector which produces the AC output.
To reduce the switching effort and guarantee zero gate losses, a special gate driver allows the switching of all bridge cells simultaneously.
Furthermore, a layered power packaging which stacks the bridges onto each other allowed a compact design achieving a power density of
400 W/in$^3$ while maintaining an efficiency of more than 99 $\%$ \cite{GBC1}. 
\newline
Inspired by those results, a second converter based on a tapped autotransformer has been introduced in \cite{GBC2}. 
\newline
Another DC/AC topology based on a multi-active bridge converter has been introduced in \cite{GBC3}. The authors introduce a multi-active bridge with increasing numbers of winding turns. Instead of a tap selector, a tap-adder is used. 
In combination with the varying number of winding turns it is possible to obtain a large number of different output voltages yielding a high resolution of the output signal and a low error in output currents.
A third DC/AC topology based on a multi-active bridge with interleaved outputs has been proposed in \cite{GBC4}. 
%%% 4  Renewable Energy Applications
Another DC/DC autotransformer topology has been presented and discussed for application in power grid connection in \cite{DDA1,DDA2,DDA3,DDA4}.
The original topology consists of several DC- power grids that are connected to an AC-line through voltage source converters
\cite{DDA1}.
Contrary to regular grid interfaces, the voltage source converters are connected with each electrically whereas the connection to the AC-grid is implemented magnetically.
The hybrid DC/DC autotransformer, an extension topology class to the DC/DC autotransformer has been suggested in
\cite{DDA2}.
Contrary to the regular DC/DC autotransformer, this topology uses a mixture of rectifier and inverter cells which yields an additional cost reduction compared to the approach presented in \cite{DDA1}.
The fault tolerance of the DC/DC autotransformer has been studied in
\cite{DDA3}.
It is shown that the DC/DC autotransformer is able to isolate DC-faults under certain stepping ratios and AC-faults under all circumstances. 
Furthermore, the authors provide guidelines on how to restore the correct operation of the converter in case one of the grids fails.
To provide experimental verification, two prototypes of the DC/DC autotransformer family were presented in
\cite{DDA4} confirming the theoretical findings of the efficiency of the topology and the results in \cite{GBC1}.
%%% 5 Other applications
Other applications of the DC-autotransformer have been explored in
\cite{Minjie1,Minjie2,Minjie3,Minjie4}.
Recently, the topology of the DC-autotransformer was furthermore applied to hard drive storage control \cite{Minjie3}, solar panels and battery voltage balancing \cite{Minjie1}, which confirmed the findings of \cite{GBC1}. 
Moreover, in \cite{Minjie4} the authors connected the topology to a multilevel converter, showing the excellent capacitor balancing capabilities of the topology.
Contrary to \cite{GBC1}, the approach of \cite{Minjie1,Minjie2,Minjie3,Minjie4} is based on a controller that can control the phase of each module at a cost of higher complexity. 
\newline
Another application of multi-active bridge converters is the usage as a power interface \cite{PowerInterface}. 
Instead of only connecting one component to the multi-active bridge, it is possible to add other components of the system to the magnetic link with additional windings. 
In this way, the authors of \cite{PowerInterface} were able to connect all high- and low power components in an electric vehicle to one single multi-active bridge interface.
% 
%%% 6. MMC for automotive
The application of multilevel converters in the automotive system has been discussed in
\cite{MMCT1,MMCT2,MMCT3}.
While state-of-the-art electric vehicles utilize conventional two-level inverters \cite{MMCT1}, several advantages of the MMC have been pointed out.
To overcome the disadvantage of large DC-link capacitors, \cite{MMCT3} suggested to connect the MMC to the battery modules directly.
A promising study on the applicability of modular multilevel converters for electric vehicles has been published in \cite{MMCT2}.
The authors showed that taking the reduced battery cost due to efficiency into account, a multilevel inverter based on Si-switches was proven to be cheaper compared to a conventional 2-level SiC converter.
\newline
% 7. This paper is structured as follows...
In this paper, we propose a centralized power management system for electric vehicles.
The proposed system is based on the DC-autotransformer and capable to regulate the complete power flow within the vehicle including the battery, motor, and low power devices while requiring only a single degree of freedom for the control.
The remainder of the paper is structured as follows:
Section 2 presents the proposed topology. In section 3 we introduce the system model. Section 4 presents the prototype construction and section 5 the results. Section 6 gives the conclusion.
\section{Proposed Topology}
\begin{figure*}[h]
\centering
\ctikzset{tripoles/spdt/height=.3}
\ctikzset{tripoles/spdt/width=.5}
\ctikzset{bipoles/battery2/width=.2}
\ctikzset{multipoles/rotary/shape=circ}
\resizebox{\linewidth}{!}{
\begin{circuitikz}[european]
%%%% HV Battery
% Box
\draw (0,0) rectangle (2,2);
% Inside 
\draw (1,1.75) to[battery] (1,0.25);
% Undertext
\node at (1,-0.3) {\textbf{HV Battery}};
%%%% DCAT
% Box
\draw (2.5,0) rectangle (6.5,2);
% Inside
\draw (3.25,1.65) node[spdt,xscale=-1] (Swu) {};
\draw (3.25,0.35) node[spdt,xscale=-1] (Swd) {};
\draw (Swu.out 2) -- (Swd.out 1);
\draw (Swu.out 1) -- ++ (-0.25,0) node[circ]{};
\draw (Swd.out 2) -- ++ (-0.25,0) node[circ]{};
\draw (Swu.out 1) ++ (-0.25,0) |-  (Swd.out 2);
\draw (Swu.in) to[cute choke, twolineschoke] (Swd.in);
\node[circ] at (3.75,1.5){};
\node at (4.75,0.2) {\small{\textbf{0V}}};
\node at (5,0.6) {\small{\textbf{100V}}};
\node at (5,1.4) {\small{\textbf{100V}}};
\node at (5,1.1) {\vdots};
\node at (4.75,1.8) {\small{\textbf{400V}}};
\node(S+)[ocirc] at (5.25,1.8) {};
\node(S-)[ocirc] at (5.25,0.2) {};
\draw (S+) -- ++ (0.75,0) node[circ]{};
\draw (S-) -- ++ (0.75,0) node[circ]{};
\draw (6,0.2) to[battery2] ++ (0,0.6);
\draw (6,1.2) to[battery2] ++ (0,0.6); 
\node[circ] at (6,0.7) {};
\node[circ] at (6,1.3) {};
\node at (6,1.1) {\vdots};
\draw (6,1.8) -- (6,1.7);
\draw (6,1.2) -- (6,1.3);
% Undertext
\node at (4.5,-0.3) {\textbf{DC-Autotransformer}};
%%%% Tap Selector
% Box
\draw (7,0) rectangle (11,2);
% Inside
 \ctikzset{tripoles/spdt/width=1.075}
 \draw (7.75,1) node[rotary switch <-=4  in 60 wiper 40,xscale=-1](S){};
 \draw (6,1.8) -- (S.out 1);
 \draw (6,0.2) -- (S.out 4);
 \draw (6,0.6875) -- (S.out 3);
 \draw (6,1.3125) -- (S.out 2);
 \node at (7.35,1.1) {\vdots};
 \draw (S.in) -- ++ (0.25,0) node[ocirc]{};
 \node at (8.8,0.2) {\textbf{0V}};
 \node at (8.8,1.8) {\textbf{400V}};
 \draw (9.1,0.2) -- (9.3,0.2);
 \draw (9.3,0.2) -- (9.3,0.6);
 \draw (9.3,0.6) -- (9.5,0.6);
 \draw (9.5,0.6) -- (9.5,1.0);
 \draw (9.5,1.0) -- (9.7,1.0);
 \draw (9.7,1.0) -- (9.7,1.4);
 \draw (9.7,1.4) -- (9.9,1.4);
 \draw (9.9,1.4) -- (9.9,1.8);
 \draw (9.9,1.8) -- (10.1,1.8);
 \draw (10.1,1.8) -- (10.1,1.4);
 \draw (10.1,1.4) -- (10.3,1.4);
 \draw (10.3,1.4) -- (10.3,1.0);
 \draw (10.3,1.0) -- (10.5,1.0);
 \draw (10.5,1.0) -- (10.5,0.6);
 \draw (10.5,0.6) -- (10.7,0.6);
 \draw (10.7,0.6) -- (10.7,0.2);
 \draw (10.7,0.2) -- (10.9,0.2);
% Undertext
\node at (9,-0.3) {\textbf{Voltage Tap Selector}};
%%%% PWM
% Box
\draw (11.5,0) rectangle (15.5,2);
% Inside
\draw (11.75,0.2) node[ocirc]{} to (12.0,0.2);
\draw (11.75,1.8) node[ocirc]{} to (12.0,1.8);
\draw (12.0,1.8) to[nos] (12.0,1.0);
\draw (12.0,1.0) to[nos] (12.0,0.2);
\node[circ] at (12.0,1.0){};
\draw (12.0,1.0) to (12.75, 1.0) node[ocirc]{};
\node at (13.1,1.0) {\textbf{AC}};
\node at (13.5,0.2) {\textbf{0V}};
\node at (13.5,1.8) {\textbf{400V}};
\draw (13.8,0.2) sin (14.6,1.8);
\draw (14.6,1.8) cos (15.3,0.2);
% Undertext
\node at (13.5,-0.3) {\textbf{Voltage Synthesizer}};
%%%% Motor
% Box
\draw (16,0) rectangle (17,2);
% Inside
 \draw (16.5,1) node[elmech](motor){M};
% Undertext
\node at (16.5,-0.3) {\textbf{Motor}};
%%%%%%%% Text on Top
\draw[>=triangle 45, <->] (1.5,2.2) -- (15.5,2.2);
\node at (0.75,2.2) {\textbf{Charing}};
\node at (16.25,2.2) {\textbf{Motoring}};
\end{circuitikz}
}
\caption{Overview of the proposed topology. The DC-autotransformer creates DC- voltage  levels. The tap selector then chooses the voltage level that achieves reference tracking. The PWM module creates a PWM-signal between the two closest taps that are on average equal to the desired output. The PWM signal is given to the motor which filters the signal.}
\label{Fig:GBC}
\end{figure*}
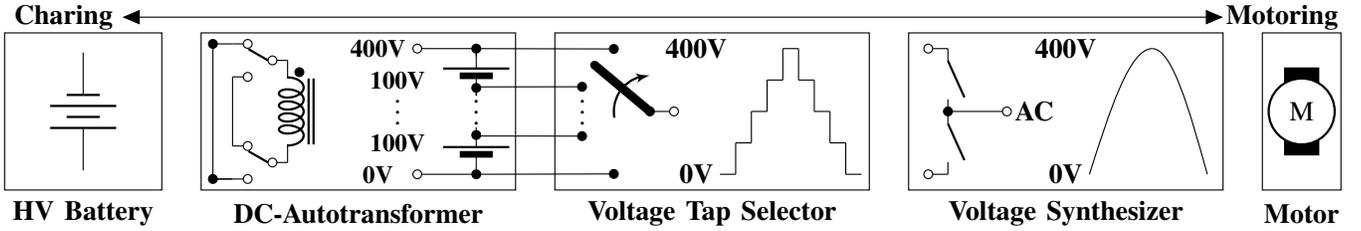
%% Complicated Circuit Diagram
\begin{figure}[!htbp]
\centering
	%% Options
	% Circuitikz
	\ctikzset{bipoles/thickness=1}
	\ctikzset{bipoles/length=1.2cm}
	\ctikzset{tripoles/thyristor/height=.8}
	\ctikzset{tripoles/thyristor/width=1}
	\ctikzset{bipoles/diode/height=.375}
	\ctikzset{bipoles/diode/width=.3}
	% Tikzstyle
	\tikzstyle{block} = [draw,fill=white, rectangle, minimum height=1cm, minimum width=6em]
	\tikzstyle{sum} = [draw, fill=white, circle, node distance=1cm]
	\tikzstyle{pinstyle} = [pin edge={to-,thin,black}]
	\resizebox{\linewidth}{!}{
    \begin{circuitikz}[european]
    %%%% Magnetic Field Lines
    \draw (-2.25,0.1) -- (-2.25,10.9);
    \draw (-2.3,0.1) -- (-2.3,10.9);
    %%% First Module
    %% Rectangles
    \draw (0,0) rectangle node{\textbf{H Bridge}} (2,2);
    % DC Voltage Ouptut
    \node at (3,1) {\textbf{100V DC}};
    % Extend to Corner
    \draw (0,0.2) -| (-0.5,0);
    \draw (0,1.8) -| (-0.5,2);
    % Extend to Trafo
    \draw(-0.5,0) -- (-2,0);
    \draw(-0.5,2) -- (-2,2);
    % Trafo Inductor
    \draw (-2,0) to[L] (-2,2);
    % Square Wave 
    \draw (-1.7,1) to (-1.7,1.5);
    \draw (-1.7,1.5) to (-1.2,1.5);
    \draw (-1.2,1.5) to (-1.2,1);
    \draw (-1.2,1) to (-0.7,1);
    \draw (-0.7,1) to (-0.7,1.5);
    % Direction Dot
    \draw[fill=black] (-1.8,1.8) circle (0.1);
    % AC Text
    \node at (-1.2,0.7){\textbf{AC}};
    % Extend to DC-source
    \draw (2,0.2) -- (4,0.2);
    \draw (2,1.8) -- (4,1.8);
    % Extend again
    \draw (4,0.2) -- (5.5,0.2);
    \draw (5.5,0.2) -- (6.5,0.2);
    % Ground it
    \node[ground] at (6.5,0.2) {};
    %%% Second Module
    %% Rectangles
    \draw (0,3) rectangle node{\textbf{H Bridge}} (2,5);
    % DC Voltage Ouptut
    \node at (3,4) {\textbf{100V DC}};
    % Extend to Corner
    \draw (0,3.2) -| (-0.5,3);
    \draw (0,4.8) -| (-0.5,5);
    % Extend to Trafo
    \draw(-0.5,3) -- (-2,3);
    \draw(-0.5,5) -- (-2,5);
    % Trafo Inductor
    \draw (-2,3) to[L] (-2,5);
    % Square Wave 
    \draw (-1.7,4) to (-1.7,4.5);
    \draw (-1.7,4.5) to (-1.2,4.5);
    \draw (-1.2,4.5) to (-1.2,4);
    \draw (-1.2,4) to (-0.7,4);
    \draw (-0.7,4) to (-0.7,4.5);
    % Direction Dot
    \draw[fill=black] (-1.8,4.8) circle (0.1);
    % AC Text
    \node at (-1.2,3.7){\textbf{AC}};
    % Extend to DC-source
    \draw (2,3.2) -- (4,3.2);
    \draw (2,4.8) -- (4,4.8);
    % Connection to current source
    \node[circ] at (4,3.2){};
    % Current Source between Modules
    \draw (4,3.2) to[american current source,l=$\mathbf{I}_1$] (4,1.8);
    % Extend again
    \draw (4,3.2) -- (6.5,3.2) node[ocirc]{};
    % Voltage Level Indication
    \node at (6.5,2.7){\textbf{100V DC}};
    % Ground it
    \node[ocirc] at (6.5,2.3){};
    \draw (6.5,2.3)-- (6.5,2.2) 
    node[ground]{};
    %%% Third Module
    %% Rectangles
    \draw (0,6) rectangle node{\textbf{H Bridge}} (2,8);
    % DC Voltage Ouptut
    \node at (3,7) {\textbf{100V DC}};
    % Extend to Corner
    \draw (0,6.2) -| (-0.5,6);
    \draw (0,7.8) -| (-0.5,8);
    % Extend to Trafo
    \draw(-0.5,6) -- (-2,6);
    \draw(-0.5,8) -- (-2,8);
    % Trafo Inductor
    \draw (-2,6) to[L] (-2,8);
    % Square Wave 
    \draw (-1.7,7) to (-1.7,7.5);
    \draw (-1.7,7.5) to (-1.2,7.5);
    \draw (-1.2,7.5) to (-1.2,7);
    \draw (-1.2,7) to (-0.7,7);
    \draw (-0.7,7) to (-0.7,7.5);
    % Direction Dot
    \draw[fill=black] (-1.8,7.8) circle (0.1);
    % AC Text
    \node at (-1.2,6.7){\textbf{AC}};
    % Extend to DC-source
    \draw (2,6.2) -- (4,6.2);
    \draw (2,7.8) -- (4,7.8);
    % Connection to current source
    \node[circ] at (4,6.2){};
    % Current Source between Modules
    \draw (4,6.2) to[american current source,l=$\mathbf{I}_2$] (4,4.8);
    % Extend again
    \draw (4,6.2) -- (6.5,6.2) node[ocirc]{};
    % Voltage Level Indication
    \node at (6.5,5.7){\textbf{200V DC}};
    % Ground it
    \node[ocirc] at (6.5,5.3){};
    \draw (6.5,5.3)-- (6.5,5.2) 
    node[ground]{};
    %%% Fourth Module
    %% Rectangles
    \draw (0,9) rectangle node{\textbf{H Bridge}} (2,11);
    % DC Voltage Ouptut
    \node at (3,10) {\textbf{100V DC}};
    % Extend to Corner
    \draw (0,9.2) -| (-0.5,9);
    \draw (0,10.8) -| (-0.5,11);
    % Extend to Trafo
    \draw(-0.5,9) -- (-2,9);
    \draw(-0.5,11) -- (-2,11);
    % Trafo Inductor
    \draw (-2,9) to[L] (-2,11);
    % Square Wave 
    \draw (-1.7,10) to (-1.7,10.5);
    \draw (-1.7,10.5) to (-1.2,10.5);
    \draw (-1.2,10.5) to (-1.2,10);
    \draw (-1.2,10) to (-0.7,10);
    \draw (-0.7,10) to (-0.7,10.5);
    % Direction Dot
    \draw[fill=black] (-1.8,10.8) circle (0.1);
    % AC Text
    \node at (-1.2,9.7){\textbf{AC}};
    % Extend to DC-source
    \draw (2,9.2) -- (4,9.2);
    \draw (2,10.8) -- (4,10.8);
    % Connection to current source
    \node[circ] at (4,9.2){};
    % Current Source between Modules
    \draw (4,9.2) to[american current source,l=$\mathbf{I}_3$] (4,7.8);
    % Extend again
    \draw (4,10.8) -- (5.5,10.8);
    \draw (5.5,10.8) -- (6.5,10.8);
    % Extend again
    \draw (4,9.2) -- (6.5,9.2) node[ocirc]{};
    % Voltage Level Indication
    \node at (6.5,8.7){\textbf{300V DC}};
    % Ground it
    \node[ocirc] at (6.5,8.3){};
    \draw (6.5,8.3)-- (6.5,8.2) 
    node[ground]{};
    %%%%% Outside Frame
    \draw (6.5,10.8) -- (9,10.8);
    \draw (6.5,0.2) -- (9,0.2);
    %% DC Source
    \draw (9,10.8) to[american voltage source, l = $\mathbf{400V}$] (9,0.2);
    \end{circuitikz}
		}
\caption{Circuit diagram of the M-active bridge converter in DC-autotransformer configuration. A DC-supply of $400$ V is connected to the highest and lowest voltage ports. In between are $4$ H-bridges in series that are connected electrically in the DC-domain and magnetically in the AC-domain. The converter produces $5$ evenly distributed output voltage levels between $0$ V and $400$ V.}
\label{Fig:Sch_Trafo}
\end{figure}
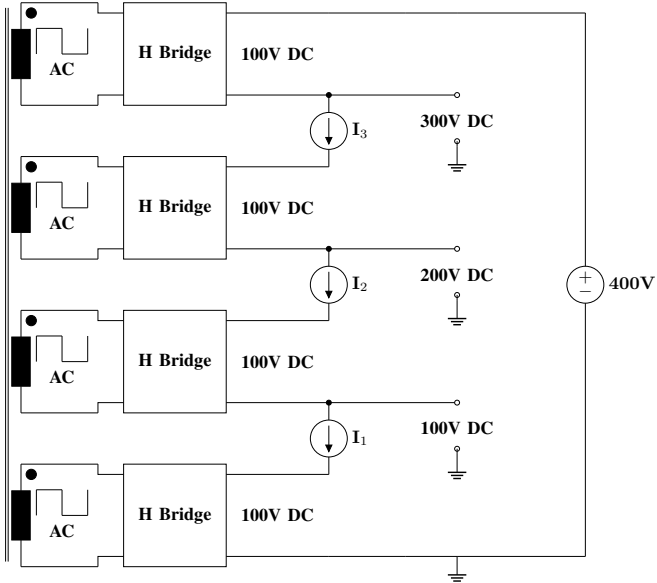
\begin{figure*}[h]
\centering
\ctikzset{tripoles/nmos/height=.7}
\ctikzset{tripoles/nmos/width=.5}
	\resizebox{\linewidth}{!}{
		\begin{tikzpicture}
	%% M - Transformer
		% Primary
        \draw (7.9,7.6)--(7.9,7.45) node (T1C1) {} to[L] ++ (-1.2,0) node (T1C2) {}  --++ (0,0.15);
        % Secondary
        \draw (13.1,7.6)--(13.1,7.45) node (T1B1) {} to[L] ++ (-1.2,0)  node (T1B2) {}  --++ (0,0.15);
        % M-ary
        \draw (19.3,7.6)--(19.3,7.45) node (T1A1) {} to[L] ++ (-1.2,0)  node (T1A2) {}  --++ (0,0.15);
        %% Iron Core
        \draw (6.7,6.97) -- (15,6.97);
        \draw (6.7,6.83) -- (15,6.83);
        \draw (16,6.97) -- (19.3,6.97);
        \draw (16,6.83) -- (19.3,6.83);
        %%% Autotransformer Nodes
        \node (j12) [circ] at (16.6,10) {};
        \node (j21) [circ] at (14.4,10) {};
        \node (j23) [circ] at (10.4,10) {};
        \node (j32) [circ] at (9.2,10) {};
        %%% Autotransformer connections
        \draw (j12) -- (16,10);
        \draw (15,10) -- (j21);
        \draw (j23) -- (j32);
		% M-ary Side MOSFETS
		\draw
		(17.35,8) node[nmos,bodydiode,rotate=270](igbt6a){}
		(19.85,8) node[nmos,bodydiode,rotate=270](igbt6b){}
		(17.35,9.5) node[nmos,bodydiode,rotate=270](igbt4a){}
		(19.85,9.5) node[nmos,bodydiode,rotate=270](igbt4b){}; 
		\draw
		%% M-ary Side Leg 4
		(16.6,9.5) to (igbt4a.S);
		\draw (igbt4a.D) to (igbt4b.S);
		\draw (igbt4b.D) to (20.6,9.5);
		% M-ary Side Leg 6
		\draw (16.6,8) coordinate (leg4) to (igbt6a.S);
		\draw (igbt6a.D) to (igbt6b.S);
		\draw (igbt6b.D) to (20.6,8) coordinate (leg6);
		% M-ary Side Horizontal Lines
		\draw (leg6) |- ++(0.5,3.0) coordinate (C5);
		\draw (leg4) |- ++(4.75,3.25) coordinate (C6);
		\draw(20.6,10.5) to[C,invert]  (16.6,10.5);
		% AC source
		%% Load
		\coordinate (V4) at (18.1,9.5);
	    \coordinate (V6) at (19.3,8);
	    % M-ary side stray inductance
		\draw (V4) node [circ] {} to (T1A2);
		\draw (V6) node [circ] {}  to[short]  (T1A1);
		%% Magnetization Inductance
		%% Secondary Side Interface with Stray Inductance
		\coordinate (V4S) at (13.1,8);
		\coordinate (V6S) at (11.9,9.5);
		\draw (V4S) node[circ]{}  to[short]  (T1B1);
		\draw (T1B2) -- (V6S) node[circ]{};
		%% Secondary Side MOSFETS
		\draw
		(11.15,8) node[nmos,bodydiode,rotate=270](igbt1a){}
		(13.65,8) node[nmos,bodydiode,rotate=270](igbt1b){}
		(11.15,9.5) node[nmos,bodydiode,rotate=270](igbt2a){}
		(13.65,9.5) node[nmos,bodydiode,rotate=270](igbt2b){}; 
		%% Secondary Leg 1
		\draw (10.4,9.5) to (igbt2a.S);
		\draw (igbt2a.D) to (igbt2b.S);
		\draw (igbt2b.D) to (14.4,9.5);
		%% Secondary Leg 2
		\draw (10.4,8) coordinate (leg2) to (igbt1a.S);
		\draw (igbt1a.D) to (igbt1b.S);
		\draw (igbt1b.D) to (14.4,8) coordinate (leg1);
		%% Horizontal Wires
		\draw (leg1)|- ++(7.45,3.75) coordinate (C3);
		\draw (leg2)|- ++(11.7,4.0) coordinate (C4);
		%% Secondary DC-Link Capacitor
		\draw (14.4,10.5) to[C]  (10.4,10.5);
		%% Secondary Side Interface with Stray Inductance
		\coordinate (V4T) at (7.9,8);
		\coordinate (V6T) at (6.7,9.5);
		\draw (V4T) node[circ]{} to[short] (T1C1);
		\draw (T1C2)to
		(V6T) node[circ]{};
		%% Primary Side MOSFETS
		\draw
		(5.95,8) node[nmos,bodydiode,rotate=270](igbt3a){}
		(8.45,8) node[nmos,bodydiode,rotate=270](igbt3b){}
		(5.95,9.5) node[nmos,bodydiode,rotate=270](igbt5a){}
		(8.45,9.5) node[nmos,bodydiode,rotate=270](igbt5b){}; 
		%% Primary Leg 5
		\draw
		(5.2,9.5) to (igbt5a.S);
		\draw (igbt5a.D) to (igbt5b.S);
		\draw (igbt5b.D) to (9.2,9.5);
		%% Primary Leg 3
		\draw (5.2,8) coordinate (leg3) to (igbt3a.S);
		\draw (igbt3a.D) to (igbt3b.S);
		\draw (igbt3b.D) to (9.2,8) coordinate (leg5);
		%% Primary Wires
		\draw (leg3) |- ++(17.4,4.5) coordinate (C1);
		\draw  (leg5) |- ++ (13.15,4.25) coordinate (C2);
		%% Primary DC-Link Capacitor
		\draw (9.2,10.5) to[C]  (5.2,10.5);
		%% Dots to show how many we have
        \node at (15.5,6.90) {$\mathbf{\hdots}$};
        \node at (15.5,10.0) {$\mathbf{\hdots}$};
        %% Connection to DC Voltage Source
        \draw(leg3) -- (5.2,6);
        \draw(leg6) -- (20.6,6);
        %% DC-Voltage Source
        \draw(5.2,6) to[battery] (20.6,6);
        %% Ground
        \node[ground] at (20.6,6){};
        %% Back Down
        \draw (C1) -- ++ (0,-2)   coordinate (D1);
        \draw (C2) -- ++ (0,-2.5) coordinate (D2);
        \draw (C4) -- ++ (0,-3)   coordinate (D4);
        \draw (C3) -- ++ (0,-3.5) coordinate (D3);
        \draw (C6) -- ++ (0,-4)   coordinate (D6);
        \draw (C5) -- ++ (0,-4.5) coordinate (D5);
        %% To the Right
        \draw (D1) -- ++ (0.50,0) node(E1)[ocirc]{};
        \draw (D2) -- ++ (0.75,0) node(E2)[ocirc]{};
        \draw (D4) -- ++ (1.00,0) node(E4)[ocirc]{};
        \draw (D3) -- ++ (1.25,0) node(E3)[ocirc]{};
        \draw (D6) -- ++ (1.75,0) node(E6)[ocirc]{};
        \draw (D5) -- ++ (2.00,0) node(E5)[ocirc]{};
        \node at (23, 7.85){\vdots};
        %% Aux
        \coordinate (F3) at (23.65,8.25);
        \coordinate (F4) at (23.65,9);
        %% Ports of TS
        \draw (23.65,8.25) node[rotary switch <-=0 in 80 wiper 20, xscale=-1] (G1) {};
        \draw (23.65,9) node[rotary switch <-=0 in 80 wiper 20, xscale=-1] (G2) {};
        %% Remaining TS
        \draw (G2.in)++(-0.05,0.05) -| ++ (0,1.5) coordinate (H2);
        \draw (G1.in)++ (-0.05,-0.05) -| ++ (0,-1.75) coordinate (H1);
        %% PWM Bridge
        \draw (H2) --++(1.5,0) coordinate (I2);
        \draw (H1) --++(1.5,0) coordinate (I1);
        \draw (I2)++(0,-1) node[nmos,bodydiode] (J2) {};
        \draw (I1)++(0,1) node[nmos,bodydiode] (J1) {};
        \draw (I1) -- (J1.S);
        \draw (J1.D) -- (J2.S);
        \draw (J2.D) -- (I2);
        \draw (I2) --++(1.25,0) coordinate (K2);
        \draw (I1) --++(1.25,0) coordinate (K1);
        \draw (K2)++(0,-1) node[nmos,bodydiode] (L2) {};
        \draw (K1)++(0,1) node[nmos,bodydiode] (L1) {};
        \draw (K1) -- (L1.S);
        \draw (L1.D) -- (L2.S);
        \draw (L2.D) -- (K2);  
        %% Output
        \draw (J2) ++ (0,-0.75) node[circ] (M2) {};
        \draw (L1) ++ (0,0.75) node[circ] (M1) {};
        \draw (M2) --++(1.75,0) coordinate (N2);
        \draw (M1) --++(0.5,0) coordinate (N1);
        \draw (N2) to node[midway,circ] (O){} (N1);
        \draw (O) to ++ (0.5,0) node (P){};
        \draw (P) to[L] ++ (1.5,0) node (Q){};
        \draw (Q) to[european resistor] ++ (0,-2.5) node[ground]{};
    	%% Boxes
		\draw[dotted] (20.9,13.0) rectangle (5,5.0); % DCAT
		\draw[dotted] (24.4,13.0) rectangle (21,5.0); % TS
		\draw[dotted] (27.4,13.0) rectangle (24.5,5.0); % PWM
		\draw[dotted] (30,13.0) rectangle (27.5,5.0); % PWM
		\end{tikzpicture}
		}
		\caption{Circuit diagram of the proposed converter. The components from left to right are the DC-autotransformer, the tap selector, the PWM module, and an $RL-$ load representing a typical motor winding. The converter is fed by a DC-source which is connected to the DC-autotransformer.}
		\label{Fig:CCD}
\end{figure*}
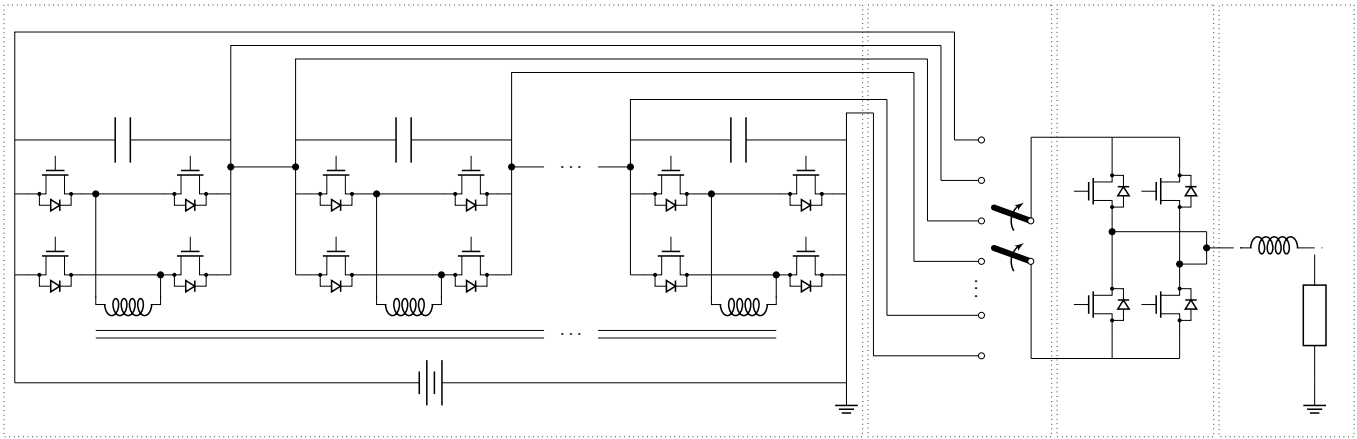
The proposed topology consists of a DC/DC converter that provides $M$ output voltage levels, a tap selector that selects one of the DC output levels, and a PWM module that generates an AC-output voltage.
An overview of the topology is given in Figure \ref{Fig:GBC}.

For the operation of the topology in electric vehicles, it is possible to directly connect the topology to the battery cells. Alternatively, the topology can as well be used as a general multilevel converter with DC-link capacitors replacing the cells.
An example of the application of the proposed topology is shown in Fig.  \ref{Fig:GBC}.
\newline
The DC/DC converter is a multi-active bridge where the bridges are connected in series in the DC-domain allowing to produce an output of the sum of the DC-voltages at all submodules.
In the AC-domain, all bridges are connected as well through a multi winding transformer ensuring a charge balance at the DC-domain.
A DC-link capacitor shall serve to buffer voltage sparks although the proposed topology requires only small DC-link capacitances.
In case full bridges are used as modules, the DC-autotransformer has $4M$ switches operating in the $100$ kHz frequency range.
The circuit Diagram of the DC-autotransformer is shown in Fig. \ref{Fig:Sch_Trafo}. \newline
To synthesize a given output voltage, a tap selector is connected to the DC-autotransformer.
The tap selector chooses two voltage levels of the DC-autotransformer and generates a PWM signal between them. 
The tap selector PWM module consists of $2$ high-frequency switches for the PWM that operate in the kHz range as well.
The tap selector itself is composed of $2M-2$ switches which operate at a frequency that is around two orders of magnitude smaller.
The whole system together with the tap selector and load is shown in Fig. \ref{Fig:CCD}.
\newline
%% Switching Losses
Since the DC-autotransformer modules operate at lower power compared to the load, it is possible to utilize more efficient low power switches for the DC-autotransformer which yields a reduction in switching losses.
The tap selector on the other hand side has to switch under high voltages and thus requires switches that can function under the full voltage of the converter.
Another limitation of the switching losses is due to the tap selector having a lower switching frequency compared to the DC-autotransformer. 
At each switching instant, the maximum number of switches of the tap selector switching is $4$ - two taps are switching off while two taps are switching on.
This number is independent of the number of modules of the circuit.
\newline
%% Conduction Losses
In addition to the low switching losses, the proposed converter has low conduction losses as well. 
It can be seen that during each switching state of the tap selector, the conduction path only consists of $2$ switches. 
The number of MOSFETs in the conduction path is furthermore independent of the number of modules contrary to many known topologies such as the modular multilevel converter. 
In addition to the conduction losses of the tap selector, the conduction losses of the DC-autotransformer can be kept low by using a low-loss planar transformer.
\newline
%% Low Energy in the system
A major feature of the proposed topology is the self-balancing effect of the DC-autotransformer. Due to their magnetic connection, the DC-output voltages of the H-bridge modules are converging to a balanced value without the need of an external controller. This advantage is particularly helpful in comparison with other multilevel inverters that only rely on electric connections and require large DC-link capacitors to absorb voltage differences due to the charging and discharging of modules during the operation.
Having smaller capacitances, the converter requires less energy to be stored in the system.
\newline
%% High Power Density
Requiring lower DC-link capacitances provides the proposed converter with high power density. 
The power density is further increased by the usage of low power switches and planar magnetics contrary to existing approaches which either require large buffer capacitors or high power switches.
\newline
%% High resolution of the system
The DC-autotransformer has a modular structure that yields a high resolution of the output signal. 
A higher resolution at the output has the advantage of providing better output signals in terms of total harmonic distortions and require less filtering.
Furthermore, a high-resolution output signal is closer to its reference which allows precise control of the tap selector.
\newline
%% Galvanic isolation
Contrary to conventional modular multilevel converters, the modules of the proposed topology are galvanically isolated. 
The galvanic isolation provides fault tolerance against failures in single modules of the converter.
In case of a module failure, the DC connection of the broken module can be bypassed and the converter still functions as an $M-1$ level converter. 
Due to the self-balancing of the H-bridges, the voltage levels are adjusted by the magnetics providing stable output levels again.
\newline
%% Simple Control
The proposed converter does not require an external balancing of the modules, which simplifies the control. 
While regular multilevel converters require complicated control structures to achieve module balancing, the proposed converter can achieve self-balancing, and an internal controller is not necessarily required. 
Thus, the major control problem lies within controlling the tap selector and the PWM module.
\newline
The advantages of the proposed topology can be summarized as
\begin{enumerate}
    \item Low conduction losses since the current only passes through one tap selector MOSFET and one PWM MOSFET for each configuration.
    \item High resolution of the AC signal due to many output levels.
    \item Low losses due to the usage of low power components.
    \item Low power inside the system due to lower DC-link capacitances.
    \item High power density due to small DC-link capacitors.
    \item Galvanic isolation of the multi-active bridge due to autotransformer connection.
    \item Simple control.
\end{enumerate}

\section{Operation of the proposed converter}
The system consists of two parts, the DC-autotransformer, and the tap selector. An advantage of the proposed converter is that both parts operate independently.

\subsection{Operation of the DC-autotransformer}
\begin{figure}[h]
    \centering
    \includegraphics[scale=0.6]{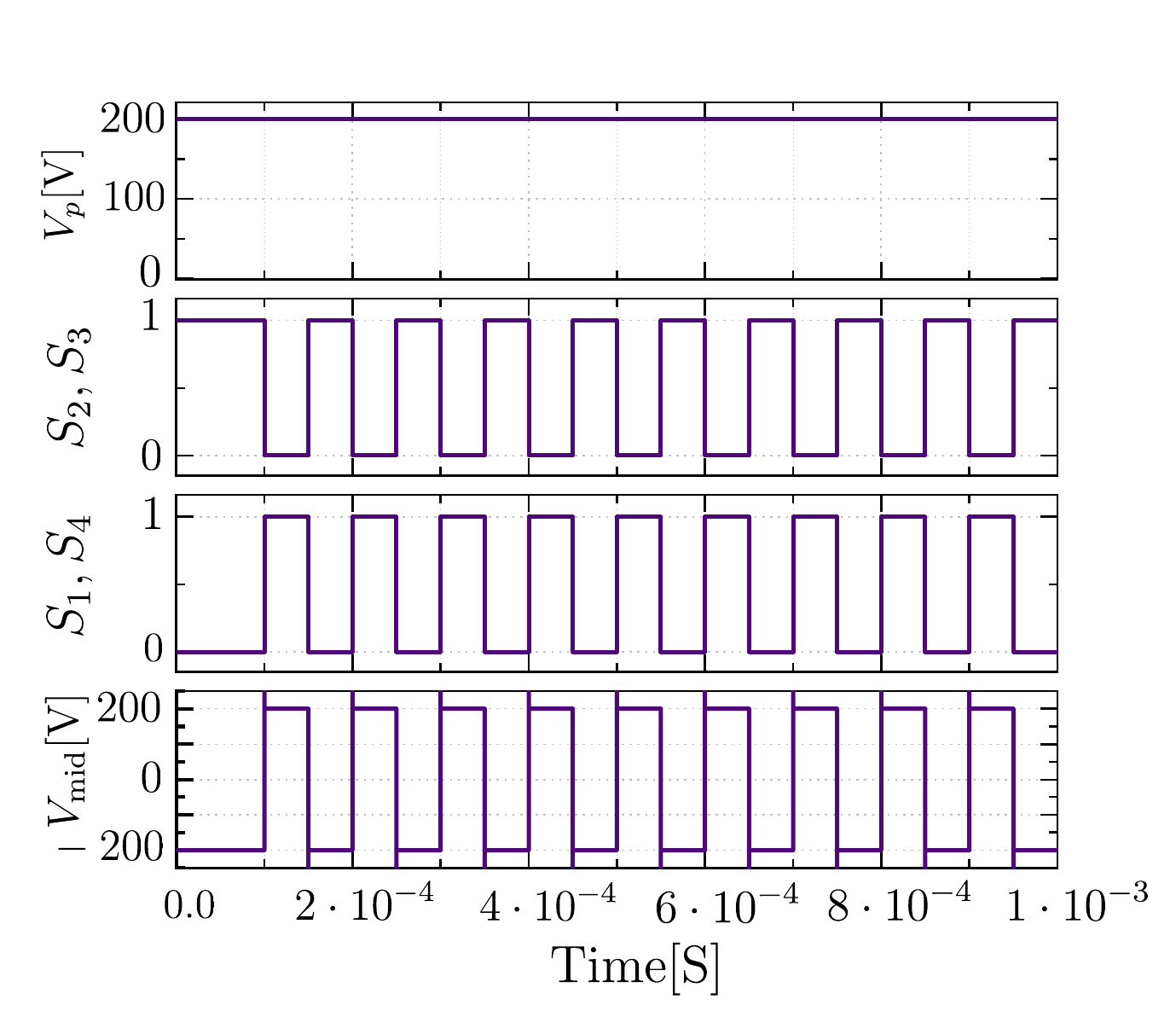}
    \caption{The functionality of an active bridge. The DC- voltage $V_p$ is given as input, $S_1$, $S_4$ and $S_2$, $S_3$ are then switching respectively forming a square-wave AC-signal at the midpoint of the bridge.}
    \label{fig:my_label}
\end{figure}
\begin{figure} [!htbp]
    \includegraphics[scale=0.27]{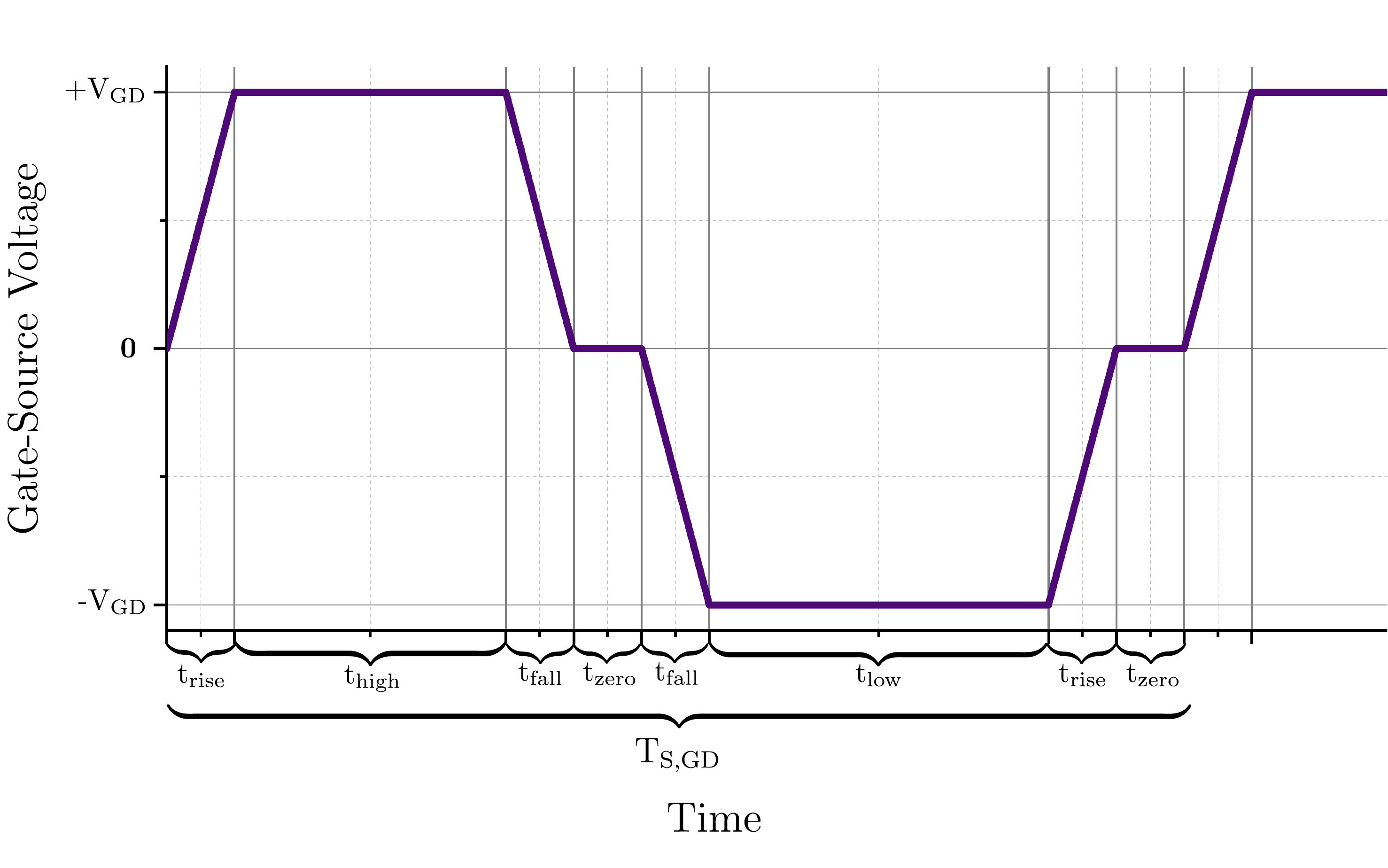}
    \caption{The desired output voltage of the gate driver board. The signal is periodic with period length $T_{\textrm{S,GD}}$. Its amplitude varies between $+V_{\textrm{GD}}$ and $-V_{\textrm{GD}}$, where $V_{\textrm{GD}}$ is the supply voltage of the gate driver. The signal itself has a piecewise linear structure where the duration of each component $t_{\textrm{rise}}$, $t_{\textrm{high}}$, $t_{\textrm{fall}}$, $t_{\textrm{zero}}$ and $t_{\textrm{low}}$ are visualized in the plot. }
    \label{Fig:GDSig}
\end{figure}
\begin{figure}[!htbp]
    \centering
    \includegraphics[scale=0.3]{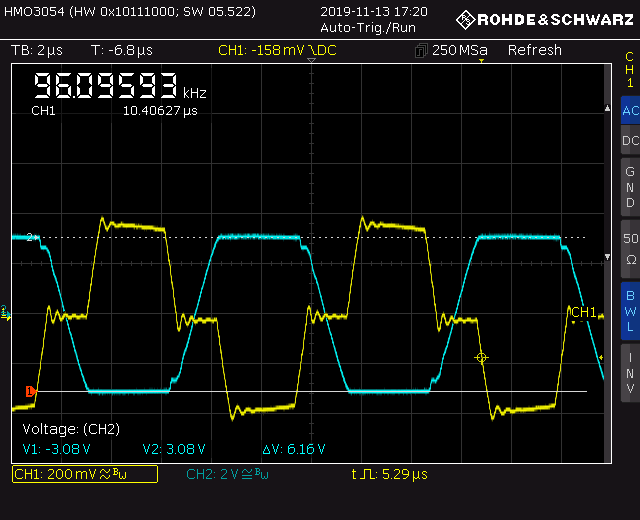}
    \caption{Waveform of the midpoint of the DC/DC converter (cyan) and the gate driver (yellow)  for a PWM frequency of $100$ kHz, optimal gate driver control and a load of $2000 \Omega$ (Experimental Results).}
    \label{Fig:SCR05}
\end{figure}

The $M$-level DC-autotransformer contains the $M$ H-bridges that generate the voltage levels.
In this paper, the functionality of the self-balancing of the converter without an external controller shall be verified. 
For this reason, each bridge is switched with a $50 \%$ duty cycle with no phase shift between the bridges.
The switching signals of one sample bridge are shown in Figure \ref{fig:my_label}.
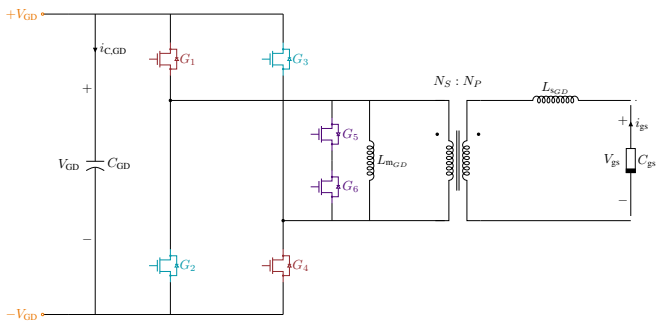
\begin{figure}[!htbp]
    \centering
	%% Options
	% Circuitikz
	\ctikzset{bipoles/thickness=1}
	\ctikzset{bipoles/length=0.8cm}
	\ctikzset{tripoles/thyristor/height=.8}
	\ctikzset{tripoles/thyristor/width=1}
	\ctikzset{bipoles/diode/height=.375}
	\ctikzset{bipoles/diode/width=.3}
	%\ctikzset{inductors/scale=0.75}
	% Tikzstyle
	\tikzstyle{block} = [draw,fill=white, rectangle, minimum height=1cm, minimum width=6em]
	\tikzstyle{sum} = [draw, fill=white, circle, node distance=1cm]
	\tikzstyle{pinstyle} = [pin edge={to-,thin,black}]
	\resizebox{\linewidth}{!}{
    \begin{tikzpicture}
      %%%%%%%%%%%%%%%%%%%%% Gate Driver %%%%%%%%%%%%%%%%%%%%%
        %% Primary Winding
		% GD Bridge MOSFETS
		\draw
		(1.7,10.5)  node[color=UCLredM,nmos,bodydiode](igbt6a){$G_{4}$}
		(1.7,16)  node[color=UCLblueB,nmos,bodydiode](igbt6b){$G_{3}$}
 		(-1.3,10.5) node[color=UCLblueB,nmos,bodydiode](igbt4a){$G_{2}$}
 		(-1.3,16) node[color=UCLredM,nmos,bodydiode](igbt4b){$G_{1}$};
		\draw
		%% GD Bridge Leg 4
		(-1.3,17.2)
		to (igbt4b.C)
		(igbt4b.E) 
		to (igbt4a.C) 
		(igbt4a.E) to (-1.3,9.2) coordinate (leg4);
		% GD Bridge Leg 6
		\draw
		(1.7,17.2)
		to (igbt6b.C);
		\draw (igbt6b.E)to(igbt6a.C); 
		\draw (igbt6a.E) to (1.7,9.2) coordinate (leg6);
		% GD Bridge Horizontal Lines
		\draw 
		(leg6) -- (-4,9.2)
		(-4,17.2) --(1.7,17.2)
		;
		\draw
		%% Rectifier
		%% DC-Link
		% GD Bridge DC-Link Capacitor
		(-3.3,17.2)
		to[pC,invert, l=$C_{\textrm{GD}}$, v>=$V_{\textrm{GD}}$, i=$i_{\textrm{C,GD}}$, current/distance=-0.8]  (-3.3,9.2)
		;
		%  GD Bridge DC-Link Sensor
		\draw
		(-4.7,17.2) node [color=UCLorange,ocirc]{} node [left] {\textcolor{UCLorange}{$+V_{\textrm{GD}}$}}  -- (-4.0,17.2)
		(-4.7,9.2) node [color=UCLorange,ocirc]{} node [left] {\textcolor{UCLorange}{$-V_{\textrm{GD}}$}}  -- (-4.0,9.2);
		% AC source
		%% Load
		\coordinate (V4) at (3.0,14.9);
	    \coordinate (V6) at (3.0,11.7);
	    \draw
		(V6)  to (leg6 |- V6) node [circ] {}
		(V4) to[short](-1.3,14.9) node [circ] {}
		;
		%% Magnetization Inductance
		\coordinate (VGDUp) at (4.0,14.9);
		\coordinate (VGDown) at (4.0,11.7);
		\coordinate (VGDmiddle) at (3.0,13.3);
		\draw
		(3.0,14.0)  node[color=UCLpurpleM,nmos,bodydiode,yscale=-1](igbtCa){}
		(3.0,12.6)  node[color=UCLpurpleM,nmos,bodydiode](igbtCb){};
		\node[right of = igbtCa, node distance=0.5cm]{\textcolor{UCLpurpleM}{$G_{5}$}};
		\node[right of = igbtCb, node distance=0.5cm]{\textcolor{UCLpurpleM}{$G_{6}$}};
	    \draw (V4) -- (igbtCa.E);
	    \draw (igbtCa.C) -- (VGDmiddle);
	    \draw (V6) -- (igbtCb.E);
	    \draw (igbtCb.C) -- (VGDmiddle);
		\draw 
		(VGDUp) to[L,l=$L_{\textrm{m}_{GD}}$,inductors/coils=7,inductors/width=1.3] (VGDown);
		%% Connection to transformer
		\draw (V6) -- ++(3.0,0) node(V6Trafo){};
		\draw (V4) -- ++(3.0,0) node(V4Trafo){};
		\node[transformer core, anchor=A1,yscale=2.6675, circuitikz/inductors/coils=7,circuitikz/inductors/width=1.3, 
		](T) at (5.75,14.9){};
		\node[circ] at (5.8,14){};
		\node[circ] at (6.9,14){};
		\draw (T.base) node (Tbase){};
		\draw node[above of=Tbase,node distance=0.5cm]{$N_S:N_P$};
        % Secondary Side
        \draw (T.B1) to[L, l=$L_{\textrm{s}_{GD}}$,inductors/coils=9,inductors/width=1.5] ++(4.0,0.0) node (GDC1){};
        \draw  (T.B2) to[short]++(4.0,0.0)node(GDC2){};
        \draw(GDC1) to[ageneric, l=$C_{\textrm{gs}}$, i<=$i_{\textrm{gs}}$,v>=$V_{\textrm{gs}}$] (GDC2);
		\end{tikzpicture}
		}
    \caption{Circuit Diagram of the Gate Driver circuit with one single winding representing the secondary side of the transformer. In the case of the real gate driver, $4M$ identical secondary windings are needed. The capacitor $C_{\textrm{gs}}$ is the gate-source capacitance of the MOSFET that the gate driver is connected to and not part of the gate driver itself. Note that some of the secondary windings are wound in different directions. The nonlinear resistor at $C_{\textrm{gs}}$ represents the nonlinear equivalent capacitance between all gates and sources.}
    \label{Fig:GD1}
\end{figure}
To achieve zero gate losses, a special gate driver is proposed. The gate driver consists of a clamped H-bridge which is connected to a transformer.
The transformer is then connected to the gates and sources of all MOSFETs in the DC-autotransformer. 
A circuit diagram of the gate driver is shown in Figure \ref{Fig:GD1} where the gate-source capacitance is replaced by one equivalent capacitor which represents all switches of the DC-autotransformer. 
The H-bridge of the gate driver is operated in the same way as the H-bridges of the DC-autotransformer.
When the voltage reaches zero, the clamped MOSFETs $5$ and $6$ are closed until the switching process of the MOSFET is finished to ensure zero gate losses.
The output signal of the gate driver is shown in Figure \ref{Fig:GDSig}.
The gate driver allows switching all MOSFETs of the gate driver at the same time. 
By using alternating winding directions, it is thus possible to achieve all switching operations of the DC-autotransformer with one single H-bridge.
The operation of the waveform was verified on the prototype and are shown in Figure \ref{Fig:SCR05}. 
It can be seen that the midpoint voltage follows the reference shown in Figure  \ref{Fig:GDSig}.

\subsection{Operation of the Tap Selector}
\begin{figure}[!htbp]
\centering
    \includegraphics[scale=0.4]{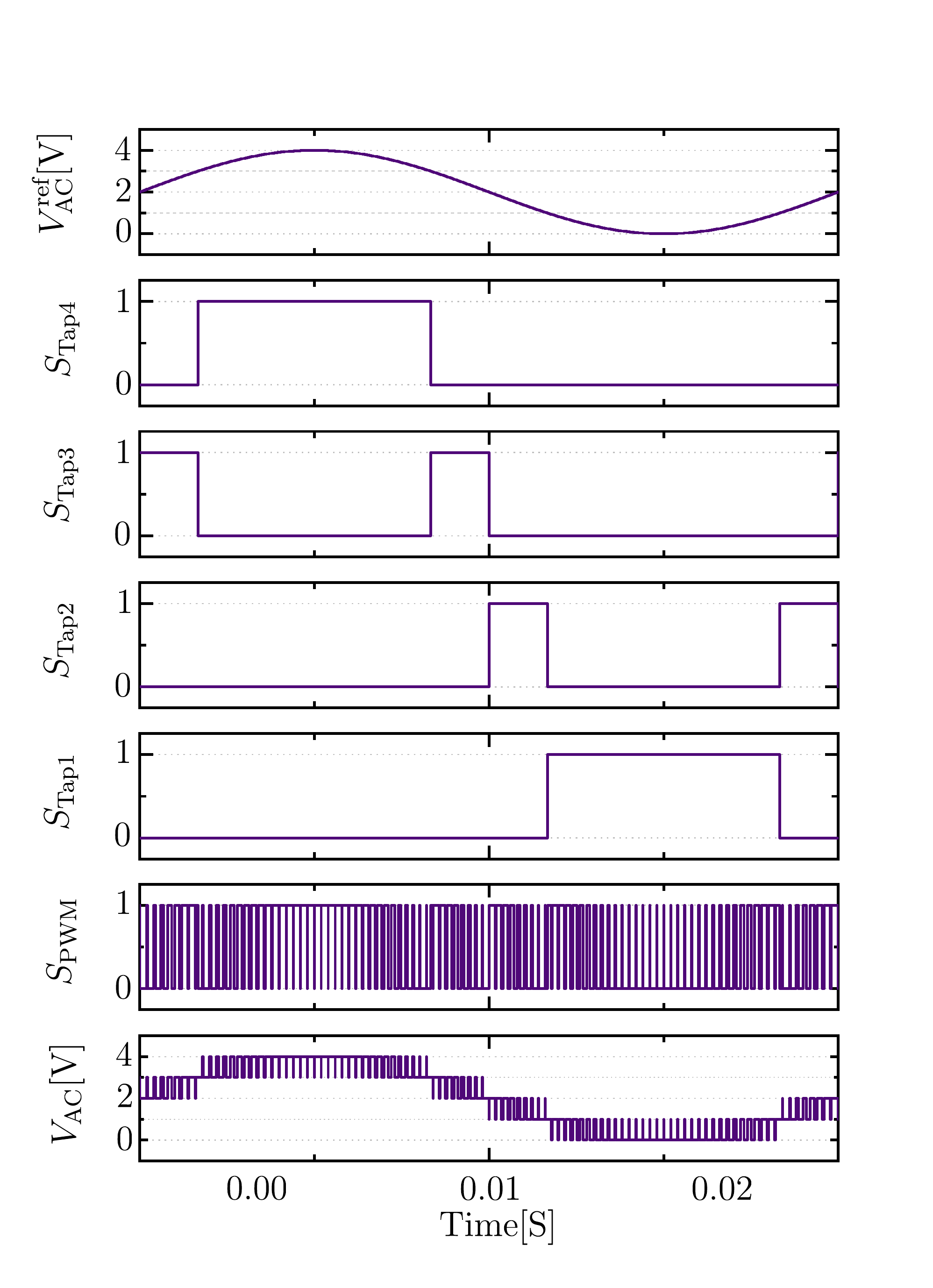}
    \caption{The functionality of the tap selector in the case of $4$ taps. Each voltage is measured with respect to the common ground of the DC-input. It is assumed that the ports are connected to a DC-autotransformer. The switching signals $S_i$ refer to the corresponding MOSFETs $G_i$. Given a reference signal, the controller produces the switching signals $S_i$ to follow it. While the signal $S_i$ determines the voltage level, the PWM-generator at the final stage of the converter determines the average voltage between the two active taps. The resulting waveform is shown in the bottom subfigure.}
    \label{Fig:TSSig}
\end{figure}

The synthesis of the AC-output signal is shown in Fig. \ref{Fig:TSSig}.
In order to follow a certain reference signal $V^{\textrm{ref}}_{\textrm{AC}}$, the tap selector always chooses the two taps which are closest to the current value of $V^{\textrm{ref}}_{\textrm{AC}}$, where
the positive tap is always selected to be higher than the reference and the negative signal is always selected to be lower compared to the reference.
The two selected voltages are then passed on to the positive and negative terminals of the PWM module.
Using duty-cycle modulation, the PWM module generates an output signal that is on average the desired reference voltage.
In case the reference output voltage is higher compared to the highest voltage that can be provided by the DC-autotransformer, the highest tap is selected and the PWM is set to $1$.
If the reference output voltage is lower compared to the lowest voltage of the DC-autotransformer, the lowest tap is selected and the PWM is set to $0$.

\section{Prototype Construction and Experiment Results}
\begin{figure}[!htbp]
    \centering
    \includegraphics[scale=0.06]{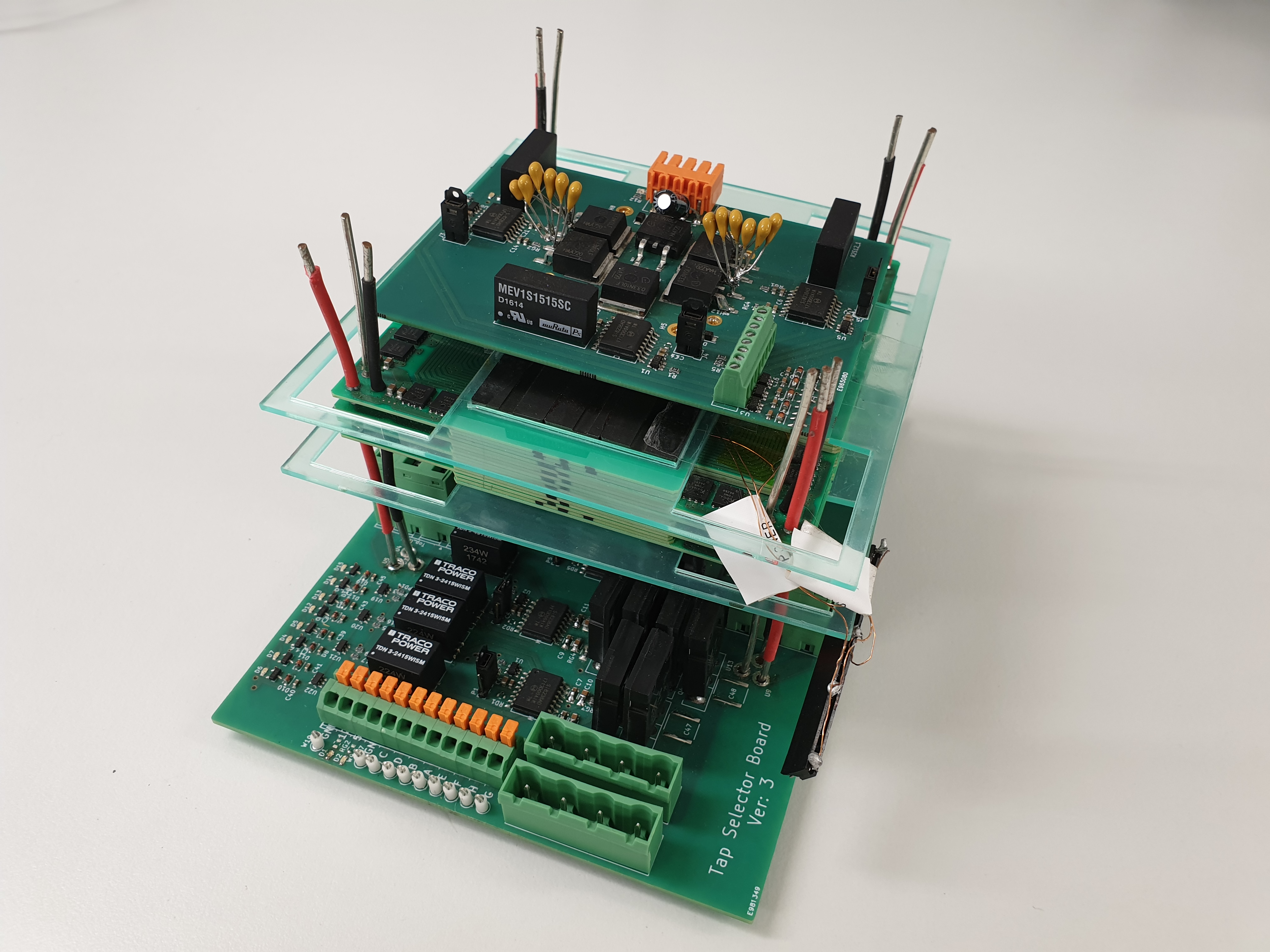}
    \caption{The assembled prototype: The gate driver is on top, the transformer board in the middle, and the tap selector on the bottom.}
    \label{Fig:Prototype}
\end{figure}

The prototype is shown in Fig. \ref{Fig:Prototype}.
The list of used components is shown in Fig. \ref{tab:Proto_Components}.
\begin{table*}[!htbp]
\caption{Choice of Components used in the Prototype}
\label{tab:Proto_Components}
\centering
\resizebox{\linewidth}{!}{%
\begin{tabular}{lllll}
\hline
Item                                       & Quantity     & Code               & Manufacturer          & Description     \\ \hline \hline 
$G_i$                                      & 4                 & IPB033N10N5LFATMA1 & Infineon Technologies & 100V Si MOSFET  \\
$G_{i,P}$, $G_{i,S}$, $G_{i,T}$, $G_{i,Q}$ & 16                & FDMS86250          & ON Semiconductor      & 150V Si MOSFET  \\
$G_{T,i}$                                  & 6                 & STY139N65M5        & STMicroelectronics    & 650V Si MOSFET  \\
$G_{PWM+}$, $G_{PWM-}$                       & 2               & STY139N65M5        & STMicroelectronics    & 650V Si MOSFET  \\ \hline
$L_m$                                      & 1         & Self Designed      & UCL                   & PCB Transformer \\
$L_s$                                      & 4                 & Self Designed      & UCL                   & PCB Transformer \\
 \\ \hline
$C_P$, $C_S$, $C_T$, $C_Q$                 &   4 (each)              & GRM31CR61H106KA12L             & Murata Electronics   &  50V Ceramic Capacitor               \\ \hline
Gate Driver               &  6        &         
SI8238AD-D-IS            & Silicon Labs  & 5kV  dual isolated gate driver               \\ \hline
\end{tabular}
}
\end{table*}
The prototype was built for a power level of $400$ V consisting of $M=4$ modules. 
The DC-autotransformer is based on a planar inductor and placed on top of the tap selector. Due to the stability of the system, each of the H-bridges of the DC-autotransformer is given a $D=0.5$ duty cycle with no phase shift between the modules. The converter sampling frequency was selected as $110$ kHz while the tap selector frequency is only $10$ kHz.
\begin{figure}[!htbp]
    \centering
    \includegraphics[scale=0.6]{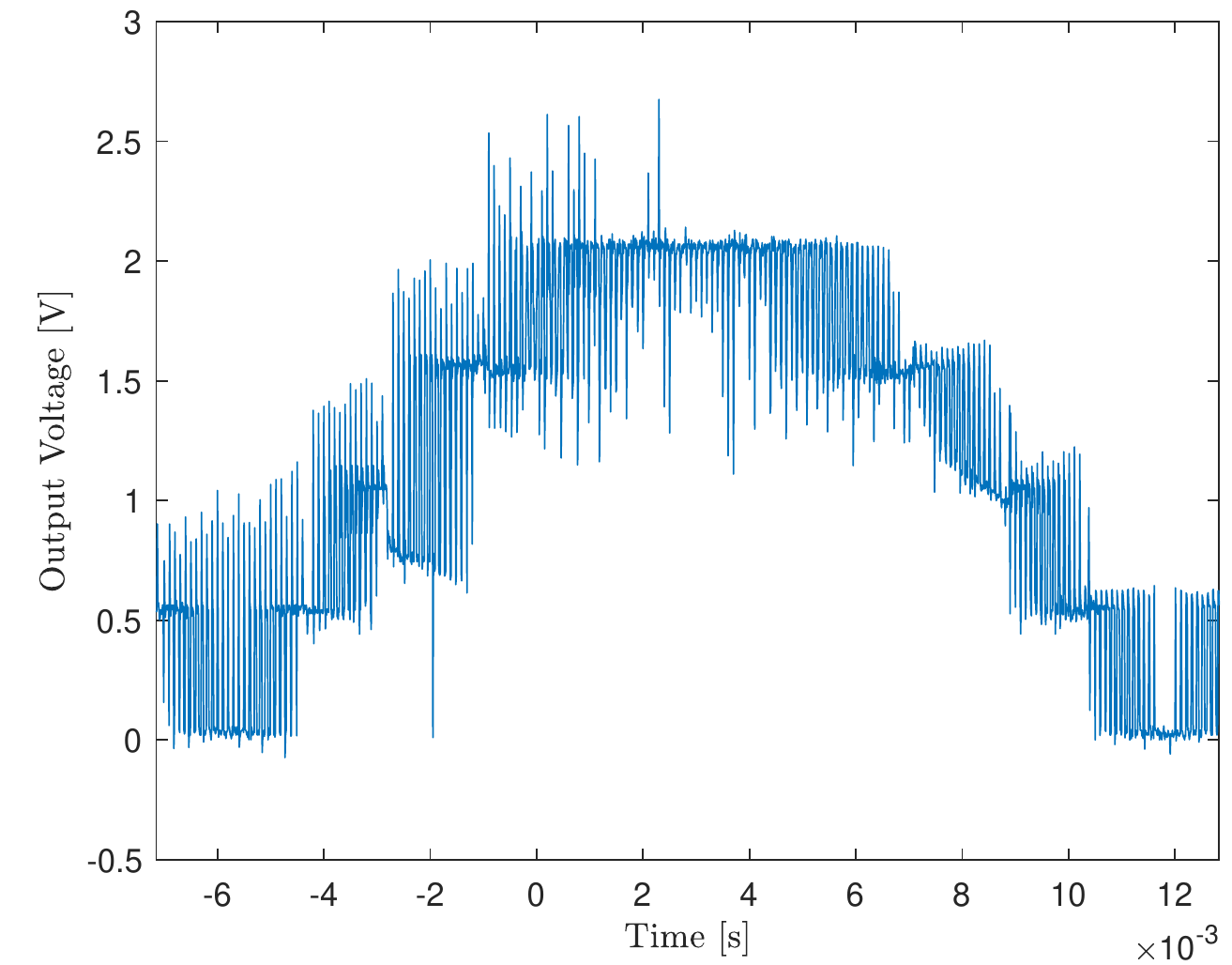}
    \caption{The output signal of the tap selector given a sine wave as reference. The voltage on the y-axis shows the one recorded by the oscilloscope through a differential probe and needs to be scaled by a factor of $10$ to obtain the real value (Experimental Results).}
    \label{Fig:TSRES}
\end{figure}
Figure \ref{Fig:TSRES} shows the signal recorded by the oscilloscope.
For the experiment, a squared sine-wave was given as reference output. 
While the signal is noisy due to being unfiltered,  the operating principle of the tap selector can be identified and similarities with Figure \ref{Fig:TSSig} are visible.

% Note that the IEEE does not put floats in the very first column
% - or typically anywhere on the first page for that matter. Also,
% in-text middle ("here") positioning is typically not used, but it
% is allowed and encouraged for Computer Society conferences (but
% not Computer Society journals). Most IEEE journals/conferences use
% top floats exclusively. 
% Note that, LaTeX2e, unlike IEEE journals/conferences, places
% footnotes above bottom floats. This can be corrected via the
% \fnbelowfloat command of the stfloats package.

\section{Conclusion}
A novel DC/AC converter for automotive applications was proposed in this paper. 
The converter consisted of a modular multilevel DC/DC converter and a tap selector. 
Using magnetic as well as electric connections in the DC/DC converter provided several advantages, including self-balancing, galvanic isolation, high efficiency, and high power density.

\appendices


% \section*{Acknowledgment}


\ifCLASSOPTIONcaptionsoff
  \newpage
\fi
\bibliographystyle{IEEEtran}
\bibliography{Paper}


%\end{IEEEbiography}


%\newpage



%\vfill


\end{document}